\documentclass[aps,prd,nofootinbib,showpacs,superscriptaddress, preprint]
{revtex4-1}
\pdfoutput=1
\usepackage[utf8]{inputenc}
\usepackage{amsmath,amssymb}
\usepackage{epsfig}
\usepackage{graphicx}
\usepackage[usenames,dvipsnames]{color}
\usepackage{comment}
\usepackage{float}
\usepackage{bbold}
\usepackage{esvect}
\usepackage{slashed}
\usepackage[colorlinks,citecolor=blue]{hyperref}
\usepackage{color}
\usepackage{subfigure}
\newcommand{\nn}{\nonumber}
\newcommand{\cen}{CE$\nu$NS~}

\begin{document}
\title{Probing Neutrino Dipole Portal at COHERENT Experiment}
\author{Jihn E. Kim}
\email{jihnekim@gmail.com}
\affiliation{Department of Physics, Kyung Hee University, 26 Gyungheedaero, Dongdaemun-Gu, Seoul 02447, Republic of Korea }
\affiliation{Department of Physics and Astronomy, Seoul National University,  1 Gwanakro, Gwanak-Gu, Seoul 08826, Republic of Korea} 
\author{Arnab Dasgupta}
\email{arnabdasgupta@protonmail.ch}
\affiliation{Institute of Convergence Fundamental Studies , Seoul National University of Science and Technology, Seoul 01811, Korea }
\author{Sin Kyu Kang}
\email{skkang@seoultech.ac.kr}
\affiliation{Institute of Convergence Fundamental Studies , Seoul National University of Science and Technology, Seoul 01811, Korea }
\affiliation{School of Liberal Arts, Seoul National University of Science and Technology, Seoul 01811, Republic of Korea}

\begin{abstract}
Motivated by the first observation of coherent-elastic neutrino-nucleus scattering  at the COHERENT experiment, we confront the neutrino dipole portal
giving rise to the transition of the standard model  neutrinos to sterile neutrinos with the recently released CENNS 10 data from the liquid argon as well as
the CsI data of the COHERENT experiment. 
Performing a statistical analysis of those data, we show how the transition magnetic moment can be constrained for the range of the sterile neutrino mass between 10 keV and 40 MeV.

\end{abstract}

\maketitle

\section{Introduction}
\label{sec:intro}
Coherent elastic neutrino-nucleus scattering (\cen) proposed roughly 47 years ago \cite{Freedman:1973yd} can be a powerful test of the Standard Model (SM), and provides a useful tool to search for new physics (NP) beyond the SM.
In 2017, the collaboration COHERENT has announced the discovery of \cen with stopped pion neutrinos on a CsI detector showing the reference of its presence against the absence at a $6.7 \sigma$ confidence level (CL) \cite{COHERENT:2017ipa,COHERENT:2018imc}.
Since then  \cen becomes a very active and dynamic field as it opens up new opportunities to probe NP \cite{Cadeddu:2017etk,Papoulias:2019lfi,Coloma:2017ncl,Liao:2017uzy,Kosmas:2017tsq,Denton:2018xmq,AristizabalSierra:2018eqm,Cadeddu:2018dux,Dutta:2019eml,Dutta:2019nbn,Cadeddu:2020lky}.
Very recently, the COHERENT has reported the observation of \cen for the first time also in argon, using a single-phase 24 kg liquid-argon (LAr) scintillation detector, with two independent analyses that prefer \cen to the background-only null hypothesis with a $3\sigma$ level \cite{COHERENT:2020iec}. 
The experimental challenge behind those experiments is the need to observe nuclear recoils with a very small kinetic energy $T_{\rm nr}$ of a few keV in the presence of a larger background.
This requirement is necessary for the coherent nucleus recoils which occur for $|\vv{q}|R <<1$ \cite{Bednyakov:2018mjd}, where $|\vv{q}|\simeq \sqrt{2m_{\mathcal{N}}T_{\rm nr}}$ is the three momentum transfer, $R$ is the nuclear radius of a few fm, and $m_{\mathcal{N}}$ is the nucleus mass.

Owing to the fact that neutrinos are massive, the SM of particle physics is incomplete.
Sterile neutrinos are natural extensions to the SM and provide a possible portal to the dark sector.
In the last few years, much attention has been paid to the models containing sterile neutrino  $N$  that can couple to the SM lepton doublet $L$ and Higgs scalar $H$. Since there is no constraint on the sterile neutrino mass,  a wide range is possible, from the sub-eV scale to the Planck scale.
This range can be somewhat narrower if $N$ is indeed taking part in generating masses for the light active neutrinos. 

In the SM, neutrinos are charge neutral and have extremely tiny electromagnetic (EM) moments \cite{Kim:2019add,Jeong:2021ivd,Kim:1976gk,Giunti:2014ixa} . 
Those EM properties of active neutrinos in the context of the SM are not detectable due to the small mass in the sub eV range with current experimental sensitivity.
However, NP beyond the SM can make them to be as large as achievable  with current sensitivity due to large transition mass at a level of keV.
It is conceivable that  new transition magnetic moments between the SM and sterile neutrinos arises, that can be larger than those predicted in the SM.
Hence the detection of such exotic EM properties would be important in probing new physics.
If observed, it implies new directions in particle physics and astrophysics.
In this paper, we attempt to address a possibility to probe the transition magnetic moments between the SM and sterile neutrinos by using \cen data.
Performing a dedicated simulation of \cen spectrum released by the COHERENT experiment on the basis of our nuclear physics calculations, we show how
the transition magnetic moment can be constrained and discuss how our results can be compared to the bounds obtained from other experiments.

The paper is organized as follows. In Sec. II, 
we briefly introduce \cen in the SM, and present how the events of \cen can be estimated and compared with
the COHERENT data. In Sec. III, we descibe the new physics interactions taken into account in this work and present formalism to study those effect on \cen.
In Sec. IV, we demonstrate how to perform $\chi^2$ fit on parameters and show the numerical results.
Finally, main conclusions are summarized in Sec. V.

\section{\cen at COHERENT and Signal Prediction}
The COHRENET experiment uses a high intensity neutrino beam produced at the Spallation Neutron Source (SNS) of the Oak Ridge National Laboratory \cite{COHERENT:2017ipa,COHERENT:2018imc,COHERENT:2020iec}. 
A 1 GeV proton beam incident on a Hg target produces pions.
The negative pions are captured almost within the target while the positive pions decay  at rest, $\pi^+  \rightarrow  \mu^+ \nu_\mu $,
produces neutrinos from which a monoenergetic  $\nu_{\mu}$ beam is generated .
The muon neutrinos arrive at the target within a short span of time $(< 1.5 \mu s)$ after  passing proton-on-target (POT) and  thus call them "prompt" neutrinos.
The muons accompanied by $\nu_{\mu}$ also decay to produce $\nu_e$ and $\bar{\nu}_\mu$ via $\mu^+ \rightarrow e^+ \nu_e \bar{\nu}_\mu $, which are known as "delayed" neutrinos and have a lower energy profile as they are produced in a three-body decay.
In the end, one can obtain  $\sim 0.08 - 0.09 $ $\nu_e$ and $\overset{\scriptscriptstyle(-)}{\nu}_{\mu}$ neutrinos per POT. 
As is well known, the neutrino fluxes from the SNS are described by,
\begin{align}
\frac{d\Phi_{\nu_\mu}}{dE_\nu} & = \eta \delta \left[E_\nu -\frac{m^2_\pi-m^2_\mu}{2m_\pi}\right], \nn \\
\frac{d\Phi_{\bar{\nu}_\mu}}{dE_\nu} & = \eta \frac{64 E^2_\nu}{m^3_\mu}\left(\frac{3}{4}-\frac{E_\nu}{m_\mu}\right), \nn \\
\frac{d\Phi_{\bar{\nu}_e}}{dE_\nu} & = \eta \frac{192 E^2_\nu}{m^3_\mu}\left(\frac{1}{2}-\frac{E_\nu}{m_\mu}\right), \nn \\
\eta &=\frac{r N_{\rm POT}}{4\pi L^2},
\end{align}
where $r$ is the number of $\nu_e$ and $\overset{\scriptscriptstyle(-)}{\nu}_{\mu}$ neutrinos produced per proton on target, $N_{\rm POT}$ is the total number of protons on target during the data-taking period and $L$ is the distance between the detector and the neutrino source.
For the dataset collected by the COHERENT CsI detector, we use $r = (8 \pm  0.9) \times 10^{-2}$, $N_{\rm POT} = 17.6 \times 10^{22}$ and $L = 19.3 {\rm m}$ \cite{COHERENT:2017ipa,COHERENT:2018imc}. 
The Ar detector, called CENNS-10 , the corresponding numbers are $r = (9 \pm  0.9) \times 10^{-2}$, $N_{\rm POT} = 13.7 \times 10^{22}$~ and $L = 27.5 {\rm m}$ \cite{COHERENT:2020iec}. 

The SM prediction for the  differential cross section of \cen  with a spin-zero nucleus  $\mathcal{N}$ with $Z$ protons and $N$ neutrons  as a function of the nuclear kinetic recoil energy $T_{\rm nr}$  is given by \cite{Drukier:1984vhf,Barranco:2005yy,Patton:2012jr}
\begin{align}
\frac{d\sigma_{\nu_l- \mathcal{N}}}{dT_{\rm nr}}(E_\nu, T_{\rm nr})=\frac{G_F^2 M_{\mathcal{N}}}{\pi}
\left( 1-\frac{M_{\mathcal{N}}T_{\rm nr}}{2E^2_\nu} \right) Q^2_{l,{\rm SM}}, \label{crossx}
\end{align}
where $G_F$ is the Fermi constant, $l=e, \mu, \tau$ denotes the neutrino flavor, $E_\nu$ is the neutrino energy and
$ Q_{l,{\rm SM}}=[g^{\rm p}_V(\nu_l) Z F_Z(|\vv{q}|^2)+g^{\rm n}_V(\nu_l) N F_N(|\vv{q}|^2)]$.
For neutrino-proton coupling, $g^{\rm p}_V$ and the neutrino-neutron coupling, $g^{\rm n}_V$,  we take the more accurate values which take into account radiative corrections in the minimal subtraction, $\overline{\rm MS}$ scheme, as follows \cite{Zyla:2020zbs};
\begin{align}
g^{\rm p}_V(\nu_e) &= 0.0401,  \\
g^{\rm n}_V(\nu_\mu) &= 0.0318, \\
g^{\rm n}_V &= 0.0401.
\end{align}

In eq. (\ref{crossx}), $F_Z(|\vv{q}|^2)$ and  $F_N(|\vv{q}|^2)$ are, respectively, the form factors of the proton and neutron distributions in the nucleus, respectively.
For those, we employ the Helm parameterisation \cite{Helm:1956zz}, that is known to be practically equivalent to the other two commonly-used symmetrized Fermi \cite{Piekarewicz:2016vbn} and Klein-Nystrand \cite{Klein:1999qj} parameterisations.
The Helm form factor is given by
\begin{align}
F_H(|\vv{q}|^2)=3\frac{j_1(|\vv{q}|R_0)}{|\vv{q}|R_0} e^{-\vv{q}^2 s^2/2},
\end{align}
where $R_p$ and $R_n$ are the rms radii of the proton and neutron distributions, respectively,
$|\vv{q}|=\sqrt{2 M_{\mathcal{N}}E_\nu}$ is the absolute value of exchanged three-momentum, $s=0.9~ {\rm fm}$
is the nuclear skin-width and $j_1(x)$ is the spherical Bessel function of order one.
The nucleon charge radius is given by
\begin{align}
R^2_{p(n)}=\frac{3}{5} R^2_0+3 s^2.
\end{align}
The size of the neutron distribution radius $R_n$ is taken to be 4.7 fm and 4.1 fm for the analyses involving CsI
and Ar, respectively \cite{Cadeddu:2020lky}.
Note that the coherence is lost for $|\vv{q}| R_{p(n)} \gtrsim 1$

The theoretical prediction of \cen event-number $N_i$ in each nuclear-recoil energy-bin $i$ is given by
\begin{align}
N_i = N(\mathcal{N})_D \int^{T_{\rm nr}^{i+1}}_{T^i_{\rm nr}} dT_{\rm nr} A(T_{\rm nr})
\int^{E_{\rm max}}_{E_{\rm min}} dE \sum_{\nu=\nu_e,\nu_\mu,\bar{\nu}_\mu}
\frac{d\Phi_\nu}{dE}\frac{d\sigma_{\nu_l- \mathcal{N}}}{dT_{\rm nr}}(E,T_{\rm nr}) \label{events}
\end{align}
where $A(T_{\rm nr})$ is the energy-dependent reconstrcution efficiency,  $E_{\rm max}=m_\mu /2 \sim 52.8$ MeV,
$N_D$ represents the number of target nuclei in the detector mass and is given by
\begin{align}
N_D = g_{\rm mol} \frac{m_{\rm det}N_A}{(M_{\mathcal{N}})_{\rm mol}},
\end{align}
where $m_{\rm det}$ is  the detector mass, $N_A$ is the Avogadro’s number, $(m_{\mathcal{N}})_{\rm mol}$ is the molar mass of the detector molecule and $g_{\rm mol}$ is the numer of atoms in a single detector molecule.
For CsI detector \cite{COHERENT:2017ipa,COHERENT:2018imc},  $m_{\rm det}=14.6 ~{\rm kg}$ and  $(M_{\rm CsI})_{\rm mol}=259.8~{\rm gram}/{\rm mol}$, and
for Ar detector \cite{COHERENT:2020iec},  $m_{\rm det}=24~{\rm kg}$ and  $(M_{\rm Ar})_{\rm mol}=39.96~{\rm gram}/{\rm mol}$.
The lower integration limit in eq. (\ref{events}) $E_{\rm min}$ is
the minimum neutrino energy required to attain a recoil energy $T_{\rm nr}$, which is given by
\begin{align}
E_{\rm min}(T_{\rm nr})=\frac{1}{2}\left[ T_{\rm nr} +\sqrt{T_{\rm nr}(T_{\rm nr}+2m_{\mathcal{N}})}\right]\left(1+\frac{M^2_{N}}{2 T_{\rm nr}m_{\mathcal{N}}}\right).
\end{align}

In Ref. \cite{COHERENT:2018imc},  the recostruction efficiency for CsI detector is given in terms of the detected number of photoelectrons $n_{\rm PE}$ by the function
\begin{align}
f(n_{\rm PE})=\frac{a}{1+\exp(-k(n_{\rm PE}-n_0))} \Theta(n_{\rm PE}-5),
\end{align}
where
\begin{align}
a&=0.6655^{+0.0212}_{-0.0384}, \nonumber \\
k&=0.4942^{+0.0335}_{-0.0131}, \nonumber \\
n_0 &=10.8507^{+0.1838}_{-0.3995}. \nonumber
\end{align}
and the function $\Theta(x)$ is defined as
\begin{align}
\Theta(x)=\left \{ \begin{array} {cc}
                           0, & x<5, \\
                          0.5, & 5\leq x<6, \\
                             1, & x\geq 6. \end{array} \right.
\end{align}
For Ar detector, we take the dector effficiency from the results in Ref. \cite{COHERENT:2020iec}.

For CsI detector, we consider the quenching factor \cite{COHERENT:2018imc},
\begin{align}
n_{\rm PE}=1.17 \left( \frac{T_{\rm nr}}{\rm keVnr}\right),
\end{align}
describing the number of photoelectrons detected by photomultiplier tubes per keV nuclear recoil energy.
It can be used to map $n_{\rm PE}$ to the recoil energy $T_{\rm nr}$ in the analysis.
For Ar detector, the electron-equivalent recoil energy $T_{\rm ee}[{\rm keV}_{\rm ee}]$, is transformed into the nuclear recoil energy \cite{COHERENT:2020iec} $T_{\rm nr}[{\rm keV}_{\rm nr}]$
thanks to the relation
\begin{align}
T_{ee}=f_Q(T_{\rm nr}) T_{\rm nr},
\end{align}
where $f_Q$ is the quenching factor, which is the ratio between the scintillation light emitted in the nuclear and electron recoils. It is  parameterised as $f_Q(T_{\rm nr}) =(0.246\pm 0.006~ {\rm keV}_{\rm nr}) + ((7.8\pm 0.9)\times 10^{-4})T_{\rm nr}$ up to 125 ${\rm{\rm keV}_{\rm nr}}$, and is kept constant for larger values. 

In this work, we consider the additional possibility that there is new physics in the neutrino sector, and examine how this may affect the signals observed by the COHERENT experiments.
\section{Neutrino Dipole Portal}
\label{sec:model}

Given the large interests in the searches of sterile neutrinos $N$, in this work we willl examine the effect of $N$ coupled to the active neutrinos via the so-called "dipole portal" encoded in the following effective Lagrangian \cite{Magill:2018jla,Shoemaker:2018vii}
\begin{align}
\mathcal{L} \supset \bar{N} (i \slashed{\partial}-M_{N})N +(\mu_{\alpha}\bar{\nu}_L^{\alpha} \sigma_{\mu \nu} F^{\mu\nu} N+h.c.), \label{dipole}
\end{align}
where $\alpha$ denotes the flavor index, $\mu_{\alpha}$ is magnetic moment, $F^{\mu\nu}$ is the electromagnetic field strength tensor and $\nu_L$ is a SM neutrino.
This interaction has been considered in the context of the MiniBooNE events \cite{Gninenko:2009ks,Gninenko:2010pr,McKeen:2010rx,Masip:2011qb,Gninenko:2012rw,Masip:2012ke,Bertuzzo:2018itn}  and has also been studied in the context of IceCube data \cite{Coloma:2019qqj} and at the upcoming SHiP experiment \cite{Magill:2018jla}. A summary of existing constraints can be found in Refs.\cite{Coloma:2019qqj,Magill:2018jla}.
Note that this is an effective Lagrangian that needs to be UV completed at energy scales not much larger than  $\Lambda \sim \mu^{-1}$. 
Here, we assume that the Yukawa interaction $LNH$ is so suppressed that the dipole term should play an enhanced role of  EM  interactions in the production and decay of $N$, and mixing between the SM neutrinos and the sterile neutrino is so negligible that 
neutrino oscillation involving sterile neutrino should be suppressed in our study.

The dipole term can admit  for a SM neutrino to up-scatter off a nucleus $\mathcal{N}$ to the sterile neutrino $N$ given as
\begin{align}
\nu_L + \mathcal{N}\rightarrow N + \mathcal{N}.
\end{align}
This up-scattering generates a distinct recoil spectrum.
We will particularly investigate the distortion of the event spectrum generated from the up-scattering process in the presence of the dipole interaction given in eq.(\ref{dipole}).
%
%

%
%
%
Without loss of generality, we assume that $\mu_{\alpha}$ are common for all flavors.
The differential corss section for the up-scattering process is given by \cite{Magill:2018jla, Harnik:2012ni,Balantekin:2013sda, Brdar:2020quo}
\begin{eqnarray}
\frac{d\sigma (\nu_L + \mathcal{N}\rightarrow N + \mathcal{N})}{dT_{\rm nr}}
&=&\alpha_{{\rm em}} \mu^2_{\nu} Z^2 F^2_N((|\vv{q}|^2)
\left[ \frac{1}{T_{\rm nr}}-\frac{1}{E_\nu}+M^2_N \frac{T_{\rm nr}-2E_\nu -m_{\mathcal{N}}}{4E^2_\nu T_{\rm nr} m_{\mathcal{N}}}+M^4_N \frac{T_{\rm nr}-m_{\mathcal{N}}}{8E^2_\nu T^2_{\rm nr} m^2_{\mathcal{N}}} \right], \nonumber \\
&& \label{em-cx}
\end{eqnarray}
where $Z$ is the number of proton,  $M_N$ is the right-handed neutrino mass, $\alpha_{{\rm em}}$ is the electromagnetic fine structure constant, $E_\nu$ is the neutrino energy.
Inserting Eqs. (\ref{crossx}, \ref{em-cx}) into Eq.(\ref{events}), we can calculate the total  number of events for neutrino-nucleus scattring.

From the kinematics, we see that the maximum possible recoil energy for a given neutrino energy $E_{\nu}$ is reached to
\begin{align}
T^{\rm max}_{\rm nr}(E_{\nu})=\frac{1}{2 E_{\nu}+m_{\mathcal{N}}}\left[  E^2_{\nu}-\frac{1}{2}M^2_N+\frac{E_{\nu}}{2 m_{\mathcal{N}}}\left(\sqrt{M^4_N-4 M^2_N m_{\mathcal{N}}(E_{\nu}+m_{\mathcal{N}})+4 E^2_{\nu}m^2_{ \mathcal{N}}}  -M^2_N \right)   \right]. \label{Tmax}
\end{align}
In our analysis, instead of applying eq.(\ref{Tmax}) to eq.(\ref{events}), we take $M_N$ so that $T^{\rm max}_{\rm nr}$ becomes larger
than $T^{i}_{\rm nr}$ for a given range of $E_{\nu}$ in eq.(\ref{events}). We found that the most conservative upper limit on $M_N$ satisfying $T^{\rm max}_{\rm nr}\gtrsim T^{i}_{\rm nr}$ is about $40$ MeV, which is taken as an upper limit for the scan of $M_N$ in our analysis.

\section{Numerical Analaysis and Results }
\label{sec:results}

\subsection{COHERENT data analysis}
\label{sec:constraint}

In the analysis based on the CsI COHERENT dataset, we perform a fit of the data by means of a least-squares function given as \cite{COHERENT:2017ipa,COHERENT:2018imc}
\begin{eqnarray}
\chi^2_{\text{CsI}}
&=&
\sum_{i=4}^{15}
\left(
\dfrac{
N_{i}^{\text{exp}}
-
\left(1+\alpha_{\text{c}}\right) N_{i}^{\mathrm{CE}\nu \mathrm{NS}}
-
\left(1+\beta_{\text{c}}\right) B_{i}
}{ \sigma_{i} }
\right)^2\\ \nonumber
&+&
\left( \dfrac{\alpha_{\text{c}}}{\sigma_{\alpha_{\text{c}}}} \right)^2
+
\left( \dfrac{\beta_{\text{c}}}{\sigma_{\beta_{\text{c}}}} \right)^2
,
\label{chi-CsI}
\end{eqnarray}
where $N_{i}^{\text{exp}}$ is the experimental event number per energy bin $i$, and $N_{i}^{\mathrm{CE}\nu \mathrm{NS}}$ and $B_{i}$
are the theoretical event number that is calculated as explained in Sections~\ref{sec:model} and the estimated number of background events, respectively, per
energy bin $i$. The parameters $\alpha_c$ and $ \beta_c$ are the nuisance parameters quntifying the systematic uncertainty of the signal rate and that of the background rate, respectively, with  the corresponding standard deviations $\sigma_{\alpha_{\text{c}}} = 0.112$ and $\sigma_{\beta_{\text{c}}} = 0.25$
\cite{COHERENT:2017ipa}.
$\sigma_{i}$ is the statistical uncertainty taken from Ref.~\cite{COHERENT:2017ipa,COHERENT:2018imc}.
Following Ref.~\cite{Cadeddu:2019eta}, we employ only the 12 energy bins from $i=4$ to $i=15$
of the COHERENT data, because they cover the recoil kinetic energy of the more recent
Chicago-3 quenching factor measurement~\cite{Collar:2019ihs}.

In the analysis based on the Ar COHERENT experiment,  we only take the data coming from the analysis A, whose range of interest of the nuclear recoil energy is [0, 120]~$\mathrm{keV}_{ee}$  with 12 energy bins of size equal to 10~$\mathrm{keV}_{ee}$. 
Following Ref.~\cite{Cadeddu:2020lky}, we perform a fit of the data by means of the least-squares function given by
\begin{eqnarray}
\chi^2_{\text{Ar}}
&=&
\sum_{i=1}^{12}
\left(
\dfrac{
N_{i}^{\text{exp}}
-
\eta_{\mathrm{CE}\nu \mathrm{NS}} N_i^{\mathrm{CE}\nu \mathrm{NS}}
-
\eta_{\mathrm{PBRN}} B_i^{\mathrm{PBRN}}
-
\eta_{\mathrm{LBRN}} B_i^{\mathrm{LBRN}}}
{\sigma_i}
\right)^2\\ \nonumber
&+&
\left( \dfrac{\eta_{\mathrm{CE}\nu \mathrm{NS}}-1}{\sigma_{\mathrm{CE}\nu \mathrm{NS}}} \right)^2
+
\left( \dfrac{\eta_{\mathrm{PBRN}}-1}{\sigma_{\mathrm{PBRN}}} \right)^2
+
\left( \dfrac{\eta_{\mathrm{LBRN}}-1}{\sigma_{\mathrm{LBRN}}} \right)^2
,
\label{chi-Ar}
\end{eqnarray}
where  $B_i^{\mathrm{PBRN}}$ and $B_i^{\mathrm{LBRN}}$ are the estimated number of prompt beam related background (PBRN) events and 
late beam related (LBRN) ones per energy bin $i$, respectively, and with
\begin{eqnarray}
\sigma_i^2 = \left( \sigma_i^{\mathrm{exp}} \right)^2 &+&
\left[ \sigma_{\mathrm{BRNES}} \left( B_i^{\mathrm{PBRN}} + B_i^{\mathrm{LBRN}}\right)\right]^2,\\
\sigma_{\mathrm{BRNES}} &=& \sqrt{\frac{0.058^2}{12}}=1.7\%,\\
\sigma_{\mathrm{CE}\nu \mathrm{NS}} &=& 13.4\%,\\
\sigma_{\mathrm{PBRN}} &=& 32\%,\\
\sigma_{\mathrm{LBRN}} &=& 100\%.
\end{eqnarray}
In Eq.~(\ref{chi-Ar}),
$\eta_{\mathrm{CE}\nu \mathrm{NS}}$, $\eta_{\mathrm{PBRN}}$ and $\eta_{\mathrm{LBRN}}$ 
are nuisance parameters which quantify,
respectively,
the systematic uncertainty of the signal rate and that of  the PBRN and LBRN background rate,
with the
corresponding standard deviations
$\sigma_{\mathrm{CE}\nu \mathrm{NS}}$, $\sigma_{\mathrm{PBRN}}$
and
$\sigma_{\mathrm{LBRN}}$.

\begin{figure}[tbp]
\includegraphics[width=8cm]{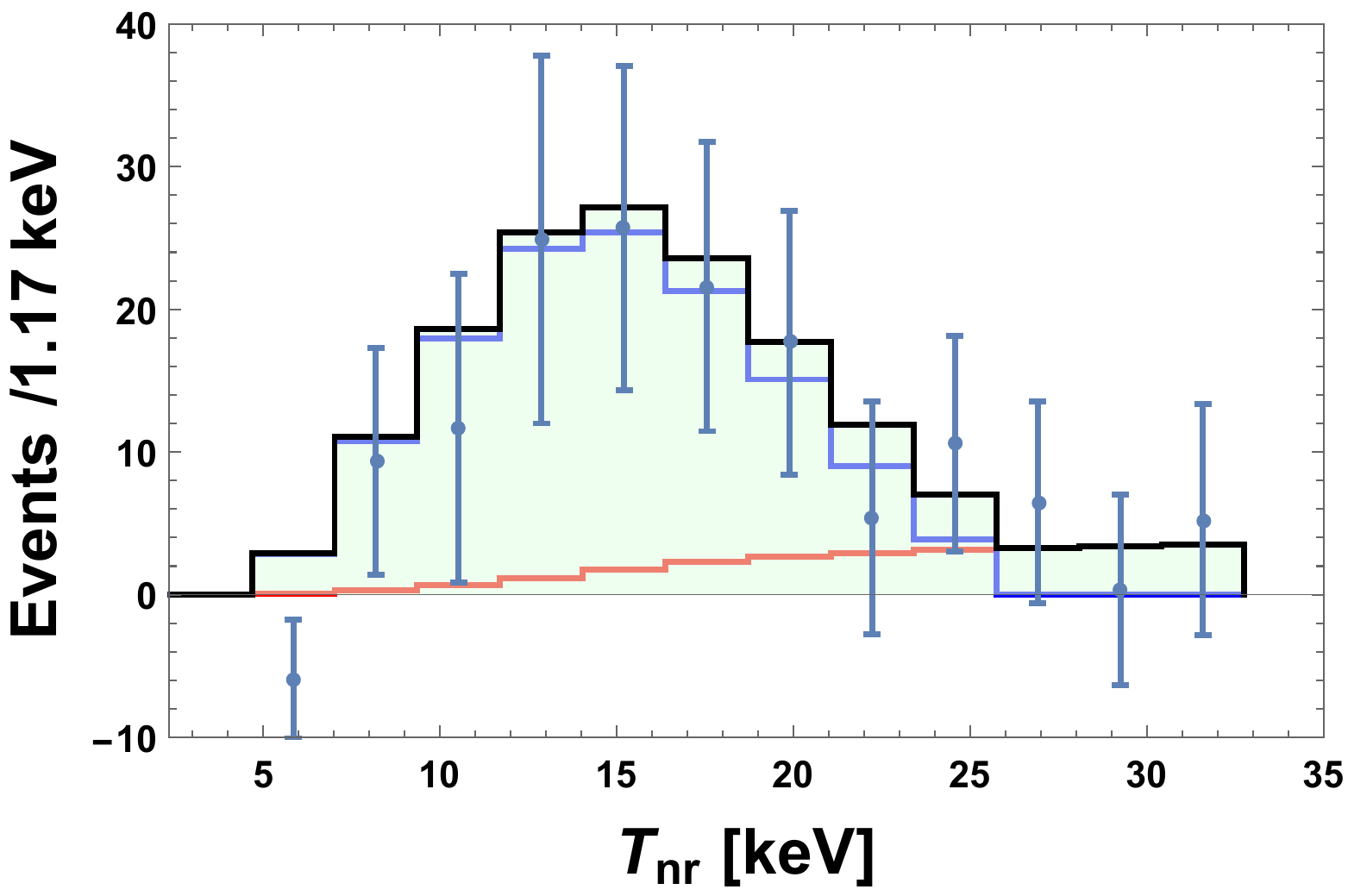}
\includegraphics[width=8cm]{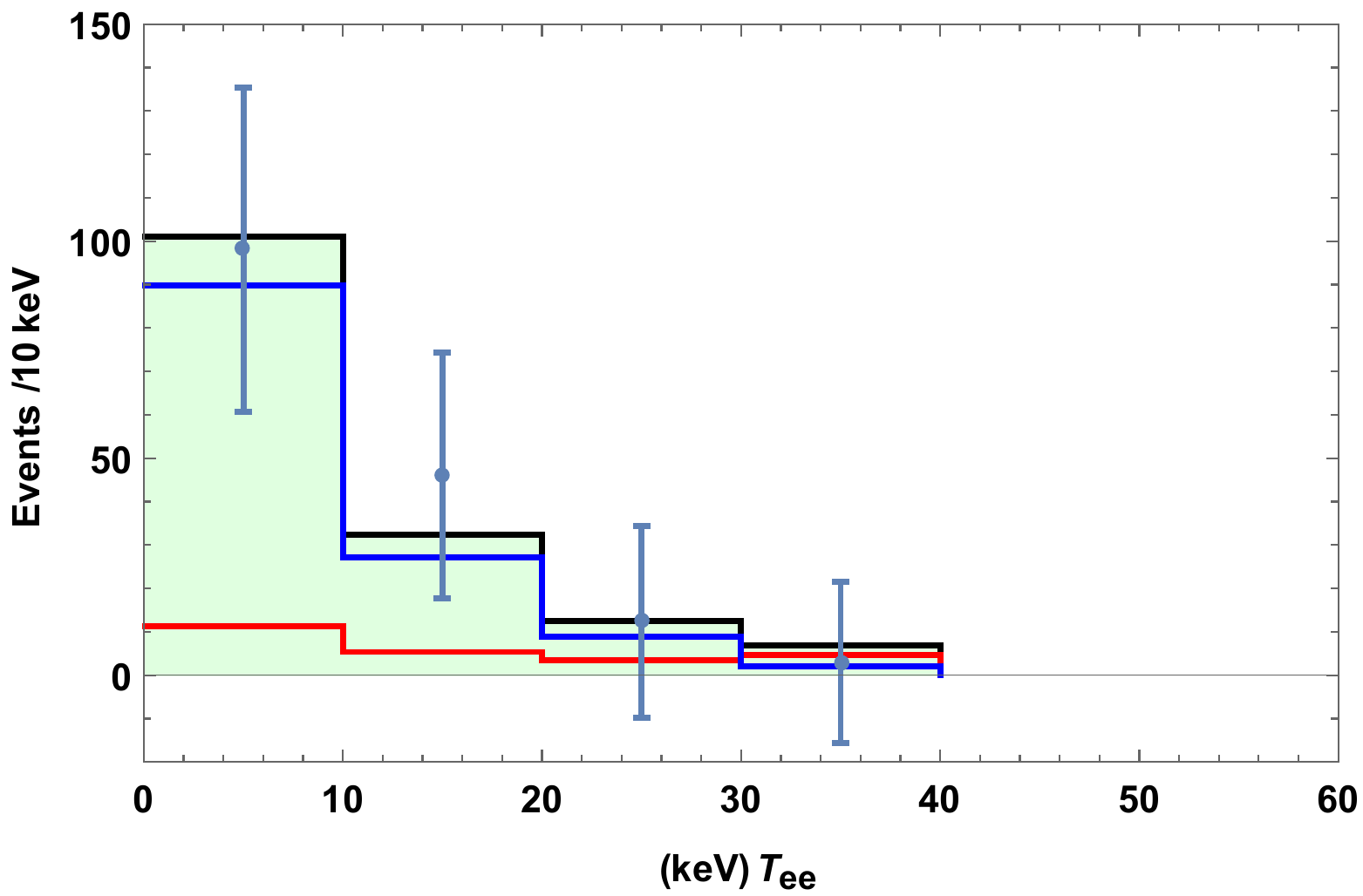}
\caption{Expected number of events for the CsI detector (Left) and Ar detector (Right)of the COHERENT collaboration in terms of
the nuclear recoil energy $T_{\rm nr}$ (keV) (Left) and $T_{\rm ee}$ (keV) (Right). The blue lines show the expected events in the SM , the red lines represent
the contribution of dipole portal to the expected number of events and the solid black line corresponds to the total number of events.
The dots with error bars correspond to the experimental measurements}\label{csi1}
\end{figure}
\begin{figure}[tbp]
\includegraphics[width=8cm]{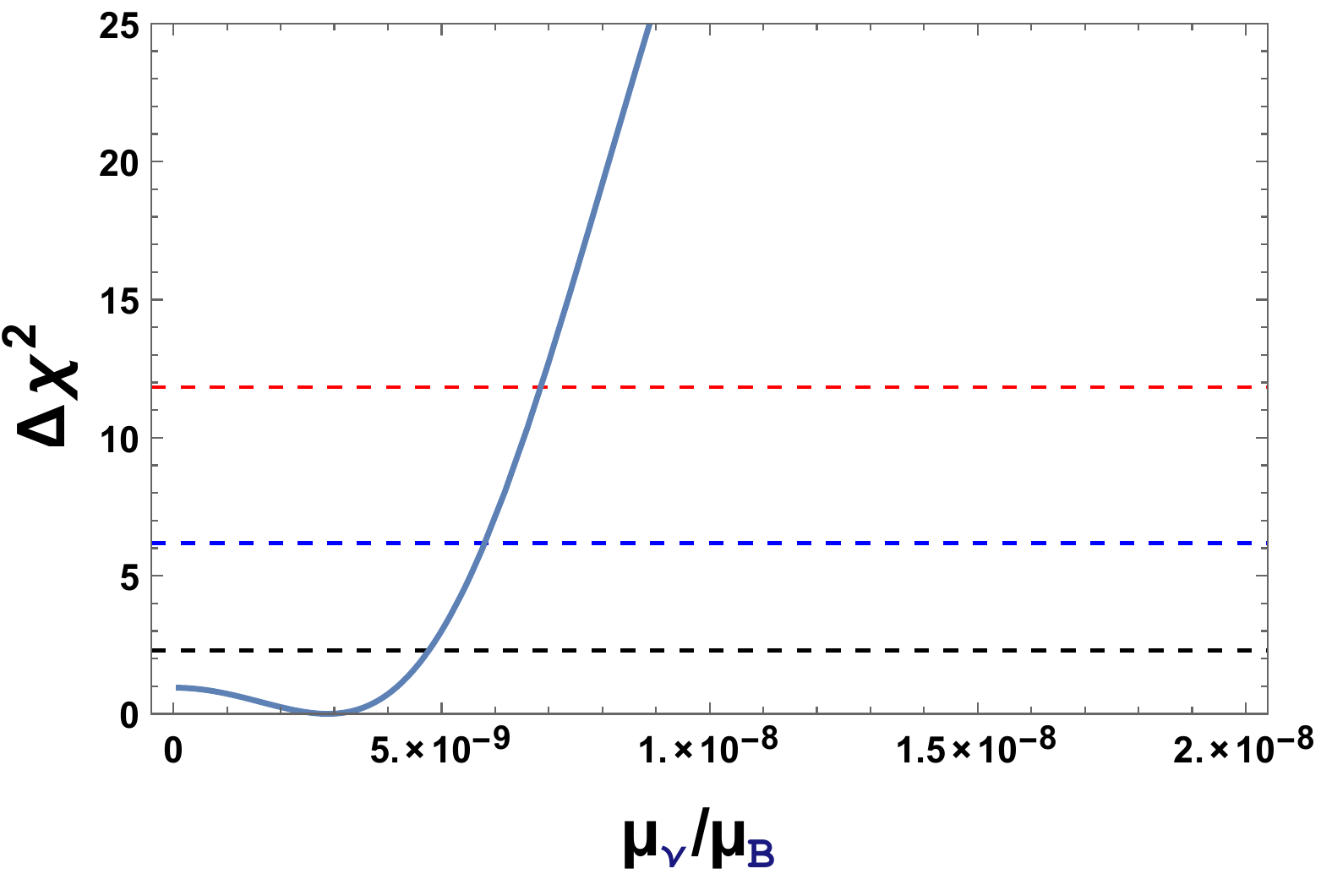}
\includegraphics[width=8cm]{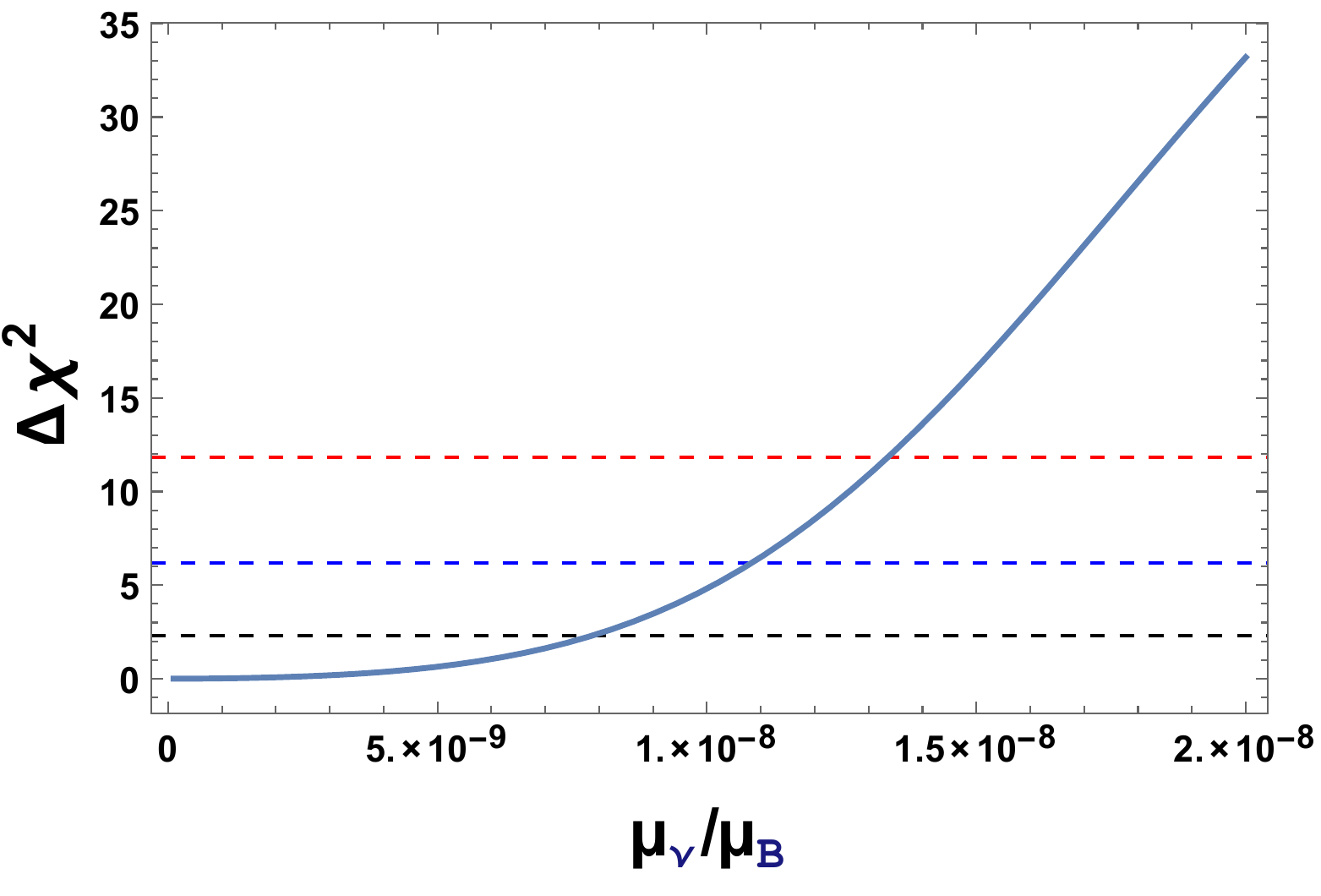}
\caption{
$\Delta \chi^2$ in terms of $\mu_{\nu}$ in unit of $\mu_B$ for CsI (Left) and Ar (Right) dector.
$M_N$ is taken as 29.8 MeV.
The black, blue and red dashed lnes correspond to the $1\sigma, 90 \%$  and $99 \%$ C.L, respectively.} \label{chi2}
\end{figure}

\begin{figure}[tbp]
\includegraphics[width=8cm]{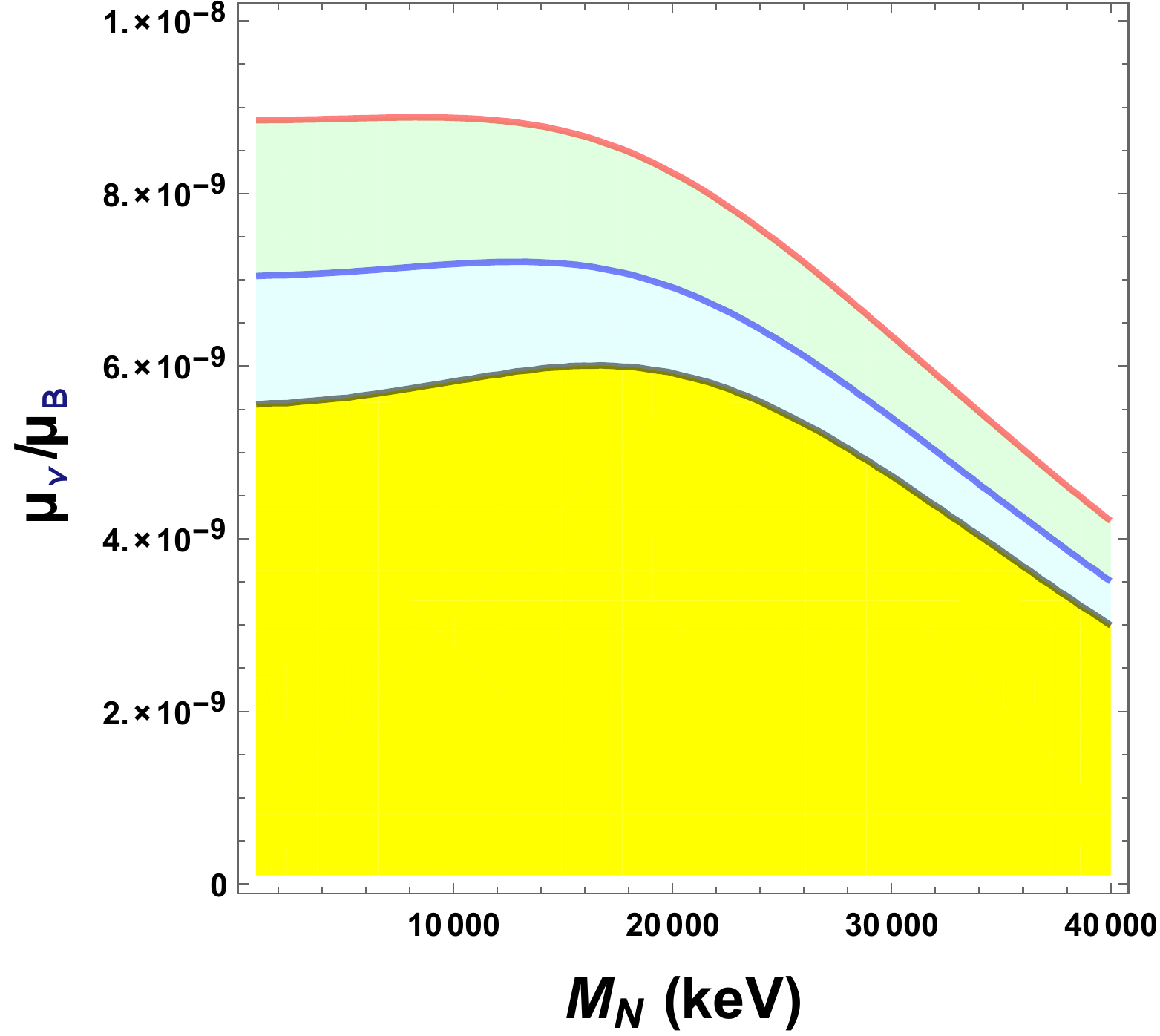}
\includegraphics[width=8cm]{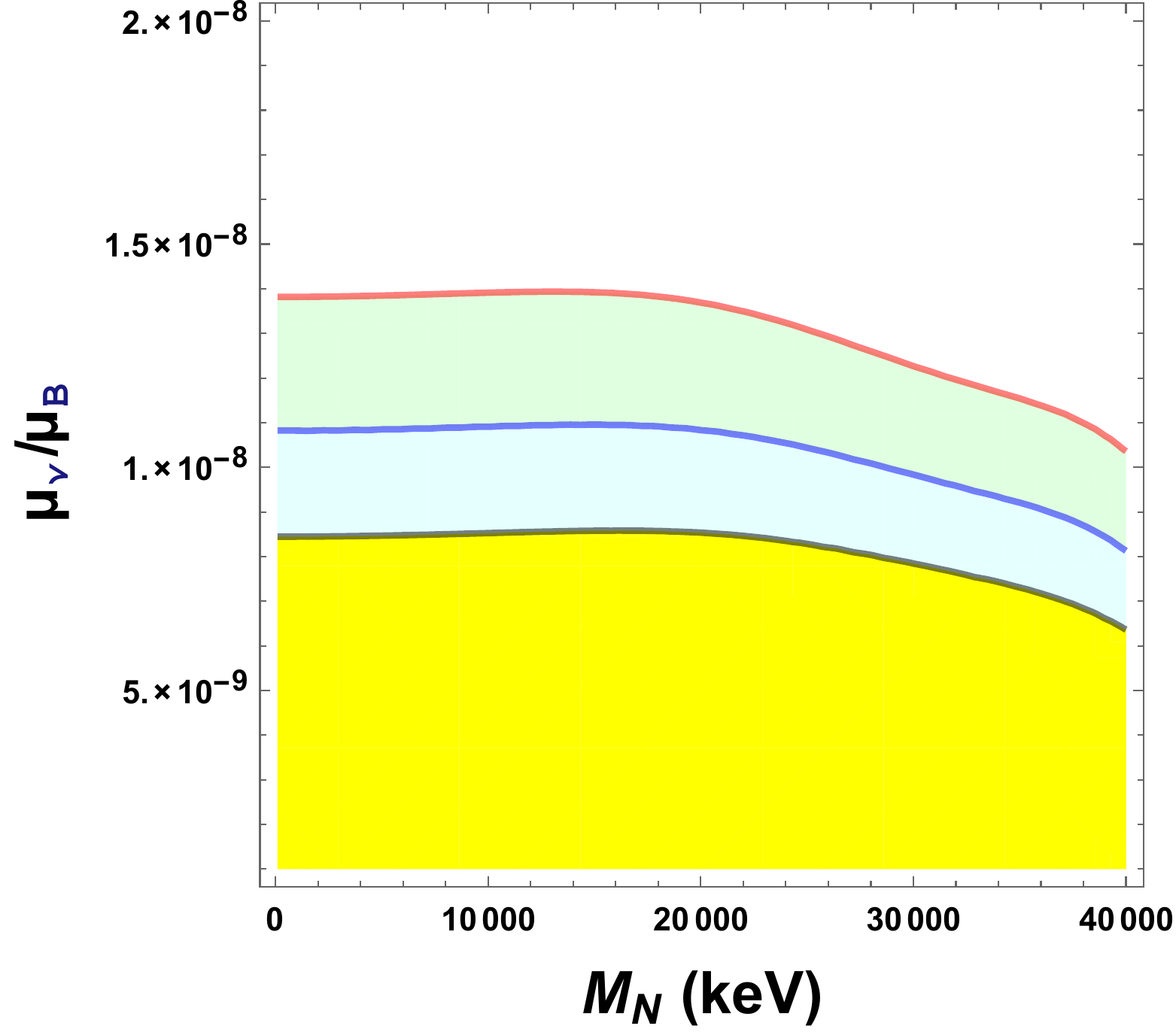}
\caption{
Allowed regions of  parameter space $(M_N, \mu_{\nu})$ for CsI (Left) and Ar (Right) dector.
The yellow regions correspond to the $1\sigma$ C.L. The light cyan and green regions are allowed regions up to $90\%$ and $99\%$ C.L., respectively.
The black, blue and red curves are the limits for $1\sigma, 90\%$ and $99\%$ C.L., respectively.
}\label{ar2}
\end{figure}

\subsection{Results and Discussion}
The parameter regions that we scan are
\begin{align}
10~{\rm keV} \lesssim M_N \lesssim 40~{\rm MeV}, \\
\mu_{\nu} \lesssim 5 \times 10^{-8} \mu_{\rm B},
\end{align}
where $ \mu_{\rm B}$ denotes Bohr magneton.
Fig.\ref{csi1} shows the expected number of events for the CsI detector (Left) and Ar detector (Right) of the COHERENT collaboration in terms of
the nuclear recoil energy $T_{\rm nr}$ (keV) (Left) and $T_{\rm ee}$ (keV) (Right), respectively.
The  left panel corresponds to the case for  $M_N=29.8$ MeV and $\mu_{\nu}/\mu_B=2.89 \times 10^{-9}$ which give the best fit of CsI data, and the right panel to the case for 
 $M_N=10$ MeV and $\mu_{\nu}/\mu_B=5 \times 10^{-9}$ taken as a bench mark point.
The blue lines show the expected events in the SM , the red  lines represent the contributions of dipole portal to the expected number of events and the solid black lines correspond to the total number of events.
The dots with error bars correspond to the experimental measurements.
We found that the best fit of the CENNS 10 data is achieved at $\mu_{\nu}=0$, which is the same as the SM result, and the minimum value of $\chi^2$ for CsI data is 6.72. It turns out that the values of $\chi^2$ for the SM predictions are 7.07 (CsI)  and 3.18 (Ar), respectively.

In Fig. \ref{chi2}, we present $\Delta \chi^2$ in terms of  $\mu_{\nu}$ in unit of $\mu_B$ for CsI (Left) and Ar (Right) dector.
The plots correpond to $M_N=28.9$ MeV.
The  black, blue and red dashed lnes correspond to the $1\sigma, 90 \%$  and $99 \%$ C.L, respectively.
From the results, we see that the minimum $\chi^2$ is located at $\mu_{\nu}\simeq 3.2 \times 10^{-9} \mu_B$ for CsI detector, while
it reaches at $\mu_{\nu}=0$ for Ar detector, which means that the SM prediction gives rise to best fit point.
The shapes of the plots in Fig. \ref{chi2} indicate that CsI data is more sensitive to the transition magnetic moment than CENNS 10 data.

Fig. \ref{ar2} shows the allowed regions of the parameter space $(M_N, \mu_{\nu})$ from the data at CsI (Left) and Ar (Right) dector.
The black, blue and red curves correspond to the limits for $ 1\sigma, 90\%$  and $99\%$ C.L., respectively.
The yellow regions correspond to the $1\sigma $ C.L. The light cyan and green regions are allowed regions up to $90\%$ and $99\%$ C.L., respectively.
For the results from CsI detector,there exist regions of parameter space $M_N$ and $\mu_{\nu}$  giving better $\chi^2$ than that for the SM prediction.
We see that for a given $M_N$ the result from the data at CsI detector gives more stringent bound on $\mu_{\nu}$ than that at Ar dector.

\begin{figure}[tbp]
\includegraphics[width=15cm]{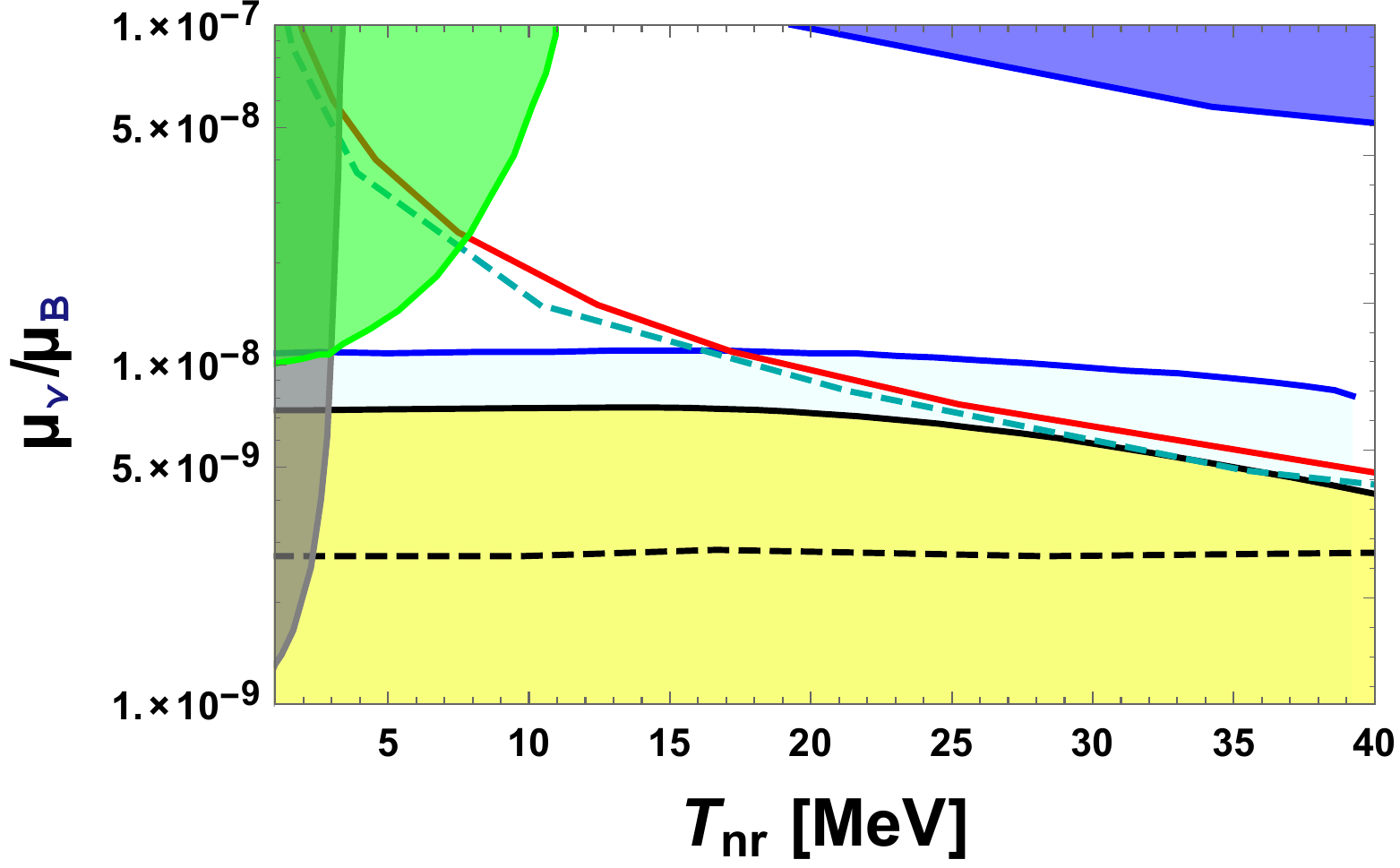}
\caption{The colored regions below the blue and black solid lines are allowed at $90\%$ by the data obtained at Ar and CsI detectors, respectively, which are the same as shown in Fig. \ref{ar2}.
The gray, green and blue regions are excluded  at $90\%$ C.L. by the data from  BOREXINO \cite{BOREXINO:2018ohr}, XENON1T  (nuclear recoil) \cite{XENON:2020rca,Shoemaker:2020kji}, and NOMAD \cite{NOMAD:1997pcg}, respectively. The red line is a part of the contour corresponding to the $90\%$ favored region at ICeCube
\cite{Coloma:2017ppo}.
The green and black dashed lines are exclusion curves obtained from MiniBooNE \cite{MiniBooNE:2007uho,Magill:2018jla} and CHARM-II \cite{CHARM-II:1989srx}.
}\label{cons}
\end{figure}

Let us compare our results with other experimental constraints shown in \cite{Shoemaker:2018vii,Brdar:2020quo}.
In Fig. \ref{cons}, the colored regions below the blue and black solid lines are allowed at $90\%$ by the data obtained at Ar and CsI detectors, respectively, which are the same as shown in Fig. \ref{ar2}.
The gray, green and blue regions are excluded  at $90\%$ C.L. by the data from  BOREXINO  \cite{BOREXINO:2018ohr}, XENON1T  (nuclear recoil) \cite{XENON:2020rca,Shoemaker:2020kji}, and NOMAD \cite{NOMAD:1997pcg}, respectively. The red line is a part of the contour corresponding to the $90\%$ favored region at ICeCube
\cite{Coloma:2017ppo}. Thus, the region above the red line is allowed.
The green and black dashed lines are exclusion curves obtained from MiniBooNE \cite{MiniBooNE:2007uho,Magill:2018jla} and CHARM-II \cite{CHARM-II:1989srx}.
Note that the constraints from the experiments except for the COHERENT correspond to the bounds on the transition magnetic moment between $\nu_{\mu}$ and $N$.

As can be seen in Fig.\ref{cons}, the result from BOREXINO shows that the bound on $\mu_{\nu}/\mu_B$ at $90\%$  for low $M_N (\lesssim 2 {\rm MeV})$ is more stringent than those we obtained in this work \cite{Shoemaker:2020kji}. 
On the other hand, for higher $M_N (\gtrsim 2 {\rm MeV})$, there is no bound on $\mu_{\nu}$ from BOREXINO  and the results of the nuclear recoil from XENON1T give the lower bound on  $\mu_{\nu}/\mu_B \sim 10^{-8}$  for  $M_N\lesssim 10~ {\rm MeV}$, whereas our results lead to new or more stringent constraints
on $\mu_{\nu}$ for $2~ {\rm MeV} \lesssim M_N \lesssim 40~ {\rm MeV}$.

Once the sterile neutrino is produced via up-scattering in the detector, it will eventually decay.
The dominant decay channel is $N\rightarrow \nu_{\alpha}+ \gamma$.
The boosted radiative decay length of the sterile neutrino with energy $E_s (=E_{\nu}-T_{\rm nr})$ is given by
\begin{align}
L_D=\gamma \beta \tau \simeq \frac{16 \pi E_s}{\mu^2_{\nu} M_N^4}\sqrt{\left(\frac{E_s}{M_N}\right)^2-1},
\end{align}
where $\gamma$ is the Lorentz factor and $\beta$ is the ratio of the velocity to speed of light in a vacuum.
For  $M_N= 5 $ MeV,  $\mu_{\nu}=5 \times 10^{-9} \mu_B$ leads to a few $m$ of $L_D$, which indicates
that the sterile neutrino decays outside the detector.
For small  $M_N (\lesssim 20 ~{\rm MeV})$,   our naive estimation shows that  the bounds on $\mu_{\nu}$ up to $99\%$ C.L. lead to rather long lived sterile neutrino decaying outside detector.

\section{Conclusion}
\label{sec:conclude}
The COHERENT collaboration has recently reported the first observation of  \cen  in CsI and liquid Ar, respectively.
Inspired that \cen can be useful to search for NP, we have exploited the experimental results in order to probe  the neutrino dipole portal
giving rise to the transition of the SM neutrinos to sterile neutrinos.
Performing a statistical analysis of the recently released CENNS 10 data from the liquid argon as well as
the CsI data,  we have examined how the transition magnetic moment between the SM neutrinos and the sterile neutrino can be constrained for the range of the sterile neutrino mass between 10 keV and 40 MeV. As a result, we found that the bounds on $\mu_{\nu}$ at $90(99)\%$
are $7.1 (89) \times 10^{-9} \lesssim \mu_{\nu}/\mu_B \lesssim 3.4 (4.2) \times 10^{-9}$ from the CsI data, whereas
$1.1 (1.4) \times 10^{-8} \lesssim \mu_{\nu}/\mu_B \lesssim 8 (10) \times 10^{-9}$ from the Ar data, which depend on $M_N$. 
The plots show that the bound on $\mu_{\nu}$ is lower as $M_N$ gets heavier.

\acknowledgments
JEK is supported in part by the NRF grant NRF-2018R1A2A3074631.  AD and SKK are supported in part by the National Research Foundation (NRF) grants NRF-2019R1A2C1088953.
%


\begin{thebibliography}{0}%
\makeatletter
\providecommand \@ifxundefined [1]{%
 \@ifx{#1\undefined}
}%
\providecommand \@ifnum [1]{%
 \ifnum #1\expandafter \@firstoftwo
 \else \expandafter \@secondoftwo
 \fi
}%
\providecommand \@ifx [1]{%
 \ifx #1\expandafter \@firstoftwo
 \else \expandafter \@secondoftwo
 \fi
}%
\providecommand \natexlab [1]{#1}%
\providecommand \enquote  [1]{``#1''}%
\providecommand \bibnamefont  [1]{#1}%
\providecommand \bibfnamefont [1]{#1}%
\providecommand \citenamefont [1]{#1}%
\providecommand \href@noop [0]{\@secondoftwo}%
\providecommand \href [0]{\begingroup \@sanitize@url \@href}%
\providecommand \@href[1]{\@@startlink{#1}\@@href}%
\providecommand \@@href[1]{\endgroup#1\@@endlink}%
\providecommand \@sanitize@url [0]{\catcode `\\12\catcode `\$12\catcode
  `\&12\catcode `\#12\catcode `\^12\catcode `\_12\catcode `\%12\relax}%
\providecommand \@@startlink[1]{}%
\providecommand \@@endlink[0]{}%
\providecommand \url  [0]{\begingroup\@sanitize@url \@url }%
\providecommand \@url [1]{\endgroup\@href {#1}{\urlprefix }}%
\providecommand \urlprefix  [0]{URL }%
\providecommand \Eprint [0]{\href }%
\providecommand \doibase [0]{http://dx.doi.org/}%
\providecommand \selectlanguage [0]{\@gobble}%
\providecommand \bibinfo  [0]{\@secondoftwo}%
\providecommand \bibfield  [0]{\@secondoftwo}%
\providecommand \translation [1]{[#1]}%
\providecommand \BibitemOpen [0]{}%
\providecommand \bibitemStop [0]{}%
\providecommand \bibitemNoStop [0]{.\EOS\space}%
\providecommand \EOS [0]{\spacefactor3000\relax}%
\providecommand \BibitemShut  [1]{\csname bibitem#1\endcsname}%
\let\auto@bib@innerbib\@empty
\end{thebibliography}%


\begin{thebibliography}{100}

\bibitem{Freedman:1973yd}
D.~Z.~Freedman,
Phys. Rev. D \textbf{9}, 1389-1392 (1974)
doi:10.1103/PhysRevD.9.1389
\bibitem{COHERENT:2017ipa}
D.~Akimov \textit{et al.} [COHERENT],
Science \textbf{357}, no.6356, 1123-1126 (2017)
doi:10.1126/science.aao0990
[arXiv:1708.01294 [nucl-ex]].
\bibitem{COHERENT:2018imc}
D.~Akimov \textit{et al.} [COHERENT],
doi:10.5281/zenodo.1228631
[arXiv:1804.09459 [nucl-ex]].
\bibitem{Cadeddu:2017etk}
M.~Cadeddu, C.~Giunti, Y.~F.~Li and Y.~Y.~Zhang,
Phys. Rev. Lett. \textbf{120}, no.7, 072501 (2018)
doi:10.1103/PhysRevLett.120.072501
[arXiv:1710.02730 [hep-ph]].
\bibitem{Papoulias:2019lfi}
D.~K.~Papoulias, T.~S.~Kosmas, R.~Sahu, V.~K.~B.~Kota and M.~Hota,
Phys. Lett. B \textbf{800}, 135133 (2020)
doi:10.1016/j.physletb.2019.135133
[arXiv:1903.03722 [hep-ph]].
\bibitem{Coloma:2017ncl}
P.~Coloma, M.~C.~Gonzalez-Garcia, M.~Maltoni and T.~Schwetz,
Phys. Rev. D \textbf{96}, no.11, 115007 (2017)
doi:10.1103/PhysRevD.96.115007
[arXiv:1708.02899 [hep-ph]].
\bibitem{Liao:2017uzy}
J.~Liao and D.~Marfatia,
Phys. Lett. B \textbf{775}, 54-57 (2017)
doi:10.1016/j.physletb.2017.10.046
[arXiv:1708.04255 [hep-ph]].
\bibitem{Kosmas:2017tsq}
D.~K.~Papoulias and T.~S.~Kosmas,
Phys. Rev. D \textbf{97}, no.3, 033003 (2018)
doi:10.1103/PhysRevD.97.033003
[arXiv:1711.09773 [hep-ph]].
\bibitem{Denton:2018xmq}
P.~B.~Denton, Y.~Farzan and I.~M.~Shoemaker,
JHEP \textbf{07}, 037 (2018)
doi:10.1007/JHEP07(2018)037
[arXiv:1804.03660 [hep-ph]].
\bibitem{AristizabalSierra:2018eqm}
D.~Aristizabal Sierra, V.~De Romeri and N.~Rojas,
Phys. Rev. D \textbf{98}, 075018 (2018)
doi:10.1103/PhysRevD.98.075018
[arXiv:1806.07424 [hep-ph]].
\bibitem{Cadeddu:2018dux}
M.~Cadeddu, C.~Giunti, K.~A.~Kouzakov, Y.~F.~Li, A.~I.~Studenikin and Y.~Y.~Zhang,
Phys. Rev. D \textbf{98}, no.11, 113010 (2018)
[erratum: Phys. Rev. D \textbf{101}, no.5, 059902 (2020)]
doi:10.1103/PhysRevD.98.113010
[arXiv:1810.05606 [hep-ph]].
\bibitem{Dutta:2019eml}
B.~Dutta, S.~Liao, S.~Sinha and L.~E.~Strigari,
Phys. Rev. Lett. \textbf{123}, no.6, 061801 (2019)
doi:10.1103/PhysRevLett.123.061801
[arXiv:1903.10666 [hep-ph]].
\bibitem{Dutta:2019nbn}
B.~Dutta, D.~Kim, S.~Liao, J.~C.~Park, S.~Shin and L.~E.~Strigari,
Phys. Rev. Lett. \textbf{124}, no.12, 121802 (2020)
doi:10.1103/PhysRevLett.124.121802
[arXiv:1906.10745 [hep-ph]].
\bibitem{Cadeddu:2020lky}
M.~Cadeddu, F.~Dordei, C.~Giunti, Y.~F.~Li, E.~Picciau and Y.~Y.~Zhang,
Phys. Rev. D \textbf{102}, no.1, 015030 (2020)
doi:10.1103/PhysRevD.102.015030
[arXiv:2005.01645 [hep-ph]].
\bibitem{COHERENT:2020iec}
D.~Akimov \textit{et al.} [COHERENT],
Phys. Rev. Lett. \textbf{126}, no.1, 012002 (2021)
doi:10.1103/PhysRevLett.126.012002
[arXiv:2003.10630 [nucl-ex]].
\bibitem{Bednyakov:2018mjd}
V.~A.~Bednyakov and D.~V.~Naumov,
Phys. Rev. D \textbf{98}, no.5, 053004 (2018)
doi:10.1103/PhysRevD.98.053004
[arXiv:1806.08768 [hep-ph]].
\bibitem{Kim:2019add}
J.~E.~Kim,
[arXiv:1911.06883 [hep-ph]].
\bibitem{Jeong:2021ivd}
J.~Jeong, J.~E.~Kim and S.~Youn,
[arXiv:2105.01842 [hep-ph]].
\bibitem{Kim:1976gk}
J.~E.~Kim,
Phys. Rev. D \textbf{14}, 3000 (1976)
doi:10.1103/PhysRevD.14.3000
\bibitem{Giunti:2014ixa}
For a review, see
C.~Giunti and A.~Studenikin,
Rev. Mod. Phys. \textbf{87}, 531 (2015)
doi:10.1103/RevModPhys.87.531
[arXiv:1403.6344 [hep-ph]].
\bibitem{Drukier:1984vhf}
A.~Drukier and L.~Stodolsky,
Phys. Rev. D \textbf{30}, 2295 (1984)
doi:10.1103/PhysRevD.30.2295
\bibitem{Barranco:2005yy}
J.~Barranco, O.~G.~Miranda and T.~I.~Rashba,
JHEP \textbf{12}, 021 (2005)
doi:10.1088/1126-6708/2005/12/021
[arXiv:hep-ph/0508299 [hep-ph]].
\bibitem{Patton:2012jr}
K.~Patton, J.~Engel, G.~C.~McLaughlin and N.~Schunck,
Phys. Rev. C \textbf{86}, 024612 (2012)
doi:10.1103/PhysRevC.86.024612
[arXiv:1207.0693 [nucl-th]].
\bibitem{Zyla:2020zbs}
P.~A.~Zyla \textit{et al.} [Particle Data Group],
PTEP \textbf{2020}, no.8, 083C01 (2020)
doi:10.1093/ptep/ptaa104
\bibitem{Helm:1956zz}
R.~H.~Helm,
Phys. Rev. \textbf{104}, 1466-1475 (1956)
doi:10.1103/PhysRev.104.1466
\bibitem{Piekarewicz:2016vbn}
J.~Piekarewicz, A.~R.~Linero, P.~Giuliani and E.~Chicken,
Phys. Rev. C \textbf{94}, no.3, 034316 (2016)
doi:10.1103/PhysRevC.94.034316
[arXiv:1604.07799 [nucl-th]].
\bibitem{Klein:1999qj}
S.~Klein and J.~Nystrand,
Phys. Rev. C \textbf{60}, 014903 (1999)
doi:10.1103/PhysRevC.60.014903
[arXiv:hep-ph/9902259 [hep-ph]].
\bibitem{Magill:2018jla}
G.~Magill, R.~Plestid, M.~Pospelov and Y.~D.~Tsai,
Phys. Rev. D \textbf{98}, no.11, 115015 (2018)
doi:10.1103/PhysRevD.98.115015
[arXiv:1803.03262 [hep-ph]].
\bibitem{Shoemaker:2018vii}
I.~M.~Shoemaker and J.~Wyenberg,
Phys. Rev. D \textbf{99}, no.7, 075010 (2019)
doi:10.1103/PhysRevD.99.075010
[arXiv:1811.12435 [hep-ph]].
\bibitem{Gninenko:2009ks}
S.~N.~Gninenko,
Phys. Rev. Lett. \textbf{103}, 241802 (2009)
doi:10.1103/PhysRevLett.103.241802
[arXiv:0902.3802 [hep-ph]].
\bibitem{Gninenko:2010pr}
S.~N.~Gninenko,
Phys. Rev. D \textbf{83}, 015015 (2011)
doi:10.1103/PhysRevD.83.015015
[arXiv:1009.5536 [hep-ph]].
\bibitem{McKeen:2010rx}
D.~McKeen and M.~Pospelov,
Phys. Rev. D \textbf{82}, 113018 (2010)
doi:10.1103/PhysRevD.82.113018
[arXiv:1011.3046 [hep-ph]].
\bibitem{Masip:2011qb}
M.~Masip and P.~Masjuan,
Phys. Rev. D \textbf{83}, 091301 (2011)
doi:10.1103/PhysRevD.83.091301
[arXiv:1103.0689 [hep-ph]].
\bibitem{Gninenko:2012rw}
S.~N.~Gninenko,
Phys. Lett. B \textbf{710}, 86-90 (2012)
doi:10.1016/j.physletb.2012.02.071
[arXiv:1201.5194 [hep-ph]].
\bibitem{Masip:2012ke}
M.~Masip, P.~Masjuan and D.~Meloni,
JHEP \textbf{01}, 106 (2013)
doi:10.1007/JHEP01(2013)106
[arXiv:1210.1519 [hep-ph]].
\bibitem{Bertuzzo:2018itn}
E.~Bertuzzo, S.~Jana, P.~A.~N.~Machado and R.~Zukanovich Funchal,
Phys. Rev. Lett. \textbf{121}, no.24, 241801 (2018)
doi:10.1103/PhysRevLett.121.241801
[arXiv:1807.09877 [hep-ph]].
\bibitem{Coloma:2019qqj}
P.~Coloma,
Eur. Phys. J. C \textbf{79}, no.9, 748 (2019)
doi:10.1140/epjc/s10052-019-7256-8
[arXiv:1906.02106 [hep-ph]].
\bibitem{Harnik:2012ni}
R.~Harnik, J.~Kopp and P.~A.~N.~Machado,
JCAP \textbf{07}, 026 (2012)
doi:10.1088/1475-7516/2012/07/026
[arXiv:1202.6073 [hep-ph]].
\bibitem{Balantekin:2013sda}
A.~B.~Balantekin and N.~Vassh,
Phys. Rev. D \textbf{89}, no.7, 073013 (2014)
doi:10.1103/PhysRevD.89.073013
[arXiv:1312.6858 [hep-ph]].

\bibitem{Brdar:2020quo}
V.~Brdar, A.~Greljo, J.~Kopp and T.~Opferkuch,
JCAP \textbf{01}, 039 (2021)
doi:10.1088/1475-7516/2021/01/039
[arXiv:2007.15563 [hep-ph]].
\bibitem{Cadeddu:2019eta}
M.~Cadeddu, F.~Dordei, C.~Giunti, Y.~F.~Li and Y.~Y.~Zhang,
Phys. Rev. D \textbf{101}, no.3, 033004 (2020)
doi:10.1103/PhysRevD.101.033004
[arXiv:1908.06045 [hep-ph]].
\bibitem{Collar:2019ihs}
J.~I.~Collar, A.~R.~L.~Kavner and C.~M.~Lewis,
Phys. Rev. D \textbf{100}, no.3, 033003 (2019)
doi:10.1103/PhysRevD.100.033003
[arXiv:1907.04828 [nucl-ex]].
\bibitem{BOREXINO:2018ohr}
M.~Agostini \textit{et al.} [BOREXINO],
Nature \textbf{562}, no.7728, 505-510 (2018)
doi:10.1038/s41586-018-0624-y
\bibitem{XENON:2020rca}
E.~Aprile \textit{et al.} [XENON],
Phys. Rev. D \textbf{102}, no.7, 072004 (2020)
doi:10.1103/PhysRevD.102.072004
[arXiv:2006.09721 [hep-ex]].
\bibitem{Shoemaker:2020kji} See aslo,
I.~M.~Shoemaker, Y.~D.~Tsai and J.~Wyenberg,
[arXiv:2007.05513 [hep-ph]].
\bibitem{NOMAD:1997pcg}
J.~Altegoer \textit{et al.} [NOMAD],
Nucl. Instrum. Meth. A \textbf{404}, 96-128 (1998)
doi:10.1016/S0168-9002(97)01079-6
\bibitem{Coloma:2017ppo}
P.~Coloma, P.~A.~N.~Machado, I.~Martinez-Soler and I.~M.~Shoemaker,
Phys. Rev. Lett. \textbf{119}, no.20, 201804 (2017)
doi:10.1103/PhysRevLett.119.201804
[arXiv:1707.08573 [hep-ph]].
\bibitem{MiniBooNE:2007uho}
A.~A.~Aguilar-Arevalo \textit{et al.} [MiniBooNE],
Phys. Rev. Lett. \textbf{98}, 231801 (2007)
doi:10.1103/PhysRevLett.98.231801
[arXiv:0704.1500 [hep-ex]].
\bibitem{CHARM-II:1989srx}
D.~Geiregat \textit{et al.} [CHARM-II],
Phys. Lett. B \textbf{232}, 539 (1989)
doi:10.1016/0370-2693(89)90457-7



\end{thebibliography}


\end{document}